\documentclass[12pt]{iopart}
\usepackage{epsf}
\def\simgt{\lower.7ex\hbox{$\;\stackrel{\textstyle>}{\sim}\;$}}
\def\simlt{\lower.7ex\hbox{$\;\stackrel{\textstyle<}{\sim}\;$}}
\begin{document}

\title[ ]{The Hypervirial-Pad\'e Summation Method Applied to the Anharmonic Oscillator}

\author{ Altu\u{g} Arda \ddag
\footnote[3]{ (arda@hacettepe.edu.tr)}\footnote[1]{This note is
dedicated to my father T. Arda.}}

\address{\ddag\
Physics Education, Hacettepe University, 06532, Beytepe, Ankara,
Turkey}

\begin{abstract}
The energy eigenvalues of the anharmonic oscillator characterized
by the cubic potential for various eigenstates are determined
within the framework of the hypervirial-Pad\'e summation method.
For this purpose the E[3,3] and E[3,4] Pad\'e approximants are
formed to the energy perturbation series and given the energy
eigenvalues up to fourth order in terms of the anharmonicity
parameter $\lambda$.

\end{abstract}

\maketitle

\section{Introduction}
There has been a great deal of interest in the analytical and
numerical investigation of the one-dimensional anharmonic
oscillator. They are of interest because of their importance in
molecular vibrations\cite{Hsue} as well as in solid state
physics\cite{Flessas,Bonham}. On the other hand, the anharmonic
oscillators with cubic and/or quartic potentials can serve as a
testing ground for the various methods based on perturbative and
nonperturbative approaches such as a group-theoretical
approach\cite{Chen1}, the multiple
scale technique\cite{Auberson}, Hill determinant method\cite{Biswas,Tater} and supersymmetric approaches\cite{Keung}. \\
\noindent In the past few decades, the hypervirial-Pad\'e
summation method are applied to various kind of
potentials\cite{Killingbeck1}, e.g. the hydrogen atom with
perturbation $\alpha r$\cite{Killingbeck2}, the quartic anharmonic
oscillator within hypervirial JWKB method\cite{Killingbeck3}, the
gaussian potential\cite{Lai1} and the screened Coulomb potential
\cite{Lai2}. In addition, the hypervirial relations and the
Hellman-Feynman theorems are applied to anharmonic
oscillators\cite{Swenson} . However, the energy perturbation
series of the anharmonic oscillator diverges asymptotically for
the perturbation parameter, so one can use the Pad\'e summation
method to recover finite results for the energy
series\cite{Loeffel}. In this note, we would like to apply the
hypervirial-Pad\'e summation method to the case of the anharmonic
oscillator with the potential
\begin{eqnarray}
V(x)=\frac{1}{2}\, \omega^2x^2+\frac{1}{2}\, \lambda
x^2+\lambda^2x^3\, ,
\end{eqnarray}
\noindent In this potential we take the cubic term as a
perturbation term, but with this term we also have a quartic
perturbation term. We pointed out that the present method can be
applied to the system described by this potential and gives the
numerical results which are in agreement with those of Ref.[17].
In Section 2, we present the formulation of the method for the
case given by Eq.(1) and a relation between the energy and the
expectation values of $x$ with various powers. With this relation
and the help of the Hellman-Feynman theorems we find an equation
which is related the energy series coefficients to the
coefficients of the power series of $\left < x^N \right
>$ and obtain recurrence relations
in powers of $\lambda$. In Section 3, we give the formula for the
energy levels up to fourth order in $\lambda$. We also evaluate
the E[3,3] and E[3,4] Pad\'e approximants to the energy series and
list the numerical results of the energy eigenvalues for the
ground and first five excited energy states in Tables. Then we
present the conclusion in Section 4.
\section{Mathematical Formulation}
The Hamiltonian for the anharmonic oscillators described by Eq.(1)
is given by
\begin{eqnarray}
H=-\frac{1}{2}\, \frac{d^2}{dx^2}+V(x)\, ,
\end{eqnarray}
\noindent where the anharmonic potential considered in this note
in terms of the perturbation parameter $\lambda$ is taken to be
\begin{eqnarray}
V(x)=\frac{1}{2}\, \omega^2x^2+\frac{1}{2}\, \lambda
x^2+\lambda^2x^3\, , \nonumber
\end{eqnarray}
\noindent Here, we use the units $m=\hbar=1$. By applying the
Hellman-Feynman theorems, one obtains the following relations
between the energy and the expectation values $\left < x^N \right
>$\cite{Swenson}
\begin{eqnarray}
E\left < x^N \right >&=&(\lambda+\omega^2)\, \frac{N+2}{2(N+1)}\,
\left < x^{N+2} \right >+\lambda^2\, \frac{2N+5}{2(N+1)}\, \left <
x^{N+3} \right > \nonumber \\ &-&\, \frac{1}{8}\, N(N-1)\, \left <
x^{N-2} \right
>\, ,
\end{eqnarray}
\noindent One can assume that the energy $E_n$ and the expectation
values $\left < x^N \right >$ can be expanded in power series of
$\lambda$ as
\begin{eqnarray}
E_n&=& \sum_{k=0}^{\infty}\, E^{(k)}_n\, \lambda^k\, , \\
\left < x^N \right >&=& \sum_{k=0}^{\infty}\, A^{(k)}_N\,
\lambda^k\, ,
\end{eqnarray}
\noindent where the energy of the unperturbed $nth$ state is
\begin{eqnarray}
E^{(0)}_n=\omega\, \left (n+\, \frac{1}{2}\right )\, ,
\end{eqnarray}
\noindent From the normalization condition that $\left < x^0
\right >=\left < 1 \right >=1$ [14], one has
\begin{eqnarray}
A^{(k)}_0=\delta_{k0}\, ,
\end{eqnarray}
\noindent The energy coefficients $E^{(k)}_n$ are related to the
coefficients $A^{(k)}_N$ trough the use of the Hellman-Feynman
theorem\cite{Killingbeck1}. From the Hellman-Feynman theorem
\begin{eqnarray}
\left < \frac{\partial V}{\partial \lambda} \right
>=\frac{\partial E}{\partial \lambda}=
\left < \frac{\partial H}{\partial \lambda} \right >\, ,
\end{eqnarray}
\noindent one can find
\begin{eqnarray}
E^{(k+1)}_n=\, \frac{1}{2(k+1)}\, A^{(k)}_2+\, \frac{2}{k+1}\,
A^{(k-1)}_3 \, \, \, \, \, \, \, k\geq 1\, ,
\end{eqnarray}
\noindent By equating the coefficients of various powers of
$\lambda$ on both sides of Eq.(3) with equations (4),(5) and (9),
we can calculate the energy coefficients $E^{(k)}_n$ in a
hierarchical manner\cite{Swenson,Lai3}. For example, we find ,
from the coefficients of $\lambda^0, \lambda^1, \lambda^2$ the
following relations
\begin{eqnarray}
A^{(0)}_N&=&\, \frac{1}{E^{(0)}_n}\Bigg[\frac{N+2}{2(N+1)}\,
\omega^2\, A^{(0)}_{N+2}-\, \frac{1}{8}\, N(N-1)\, A^{(0)}_{N-2}
\Bigg]\, , \\
A^{(1)}_N&=&\, \frac{1}{E^{(0)}_n}\Bigg[\frac{N+2}{2(N+1)}\,
\omega^2\, A^{(1)}_{N+2}+\frac{N+2}{2(N+1)}\, A^{(0)}_{N+2}-\,
\frac{1}{8}\, N(N-1)\, A^{(1)}_{N-2} \nonumber
\\ &-&E^{(1)}_nA^{(0)}_N \Bigg]\,
, \\
A^{(2)}_N&=&\, \frac{1}{E^{(0)}_n}\Bigg[\frac{N+2}{2(N+1)}\,
\omega^2\, A^{(2)}_{N+2}+\frac{N+2}{2(N+1)}\, A^{(1)}_{N+2}+\,
\frac{2N+5}{2(N+1)}\, A^{(0)}_{N+3} \nonumber \\&-&\,
\frac{1}{8}\, N(N-1)\,
A^{(2)}_{N-2}-E^{(1)}_nA^{(1)}_N-E^{(2)}_nA^{(0)}_N \Bigg]\, ,
\end{eqnarray}
\noindent From the above relations one can calculate the energy
coefficients $E^{(k)}_n$ from the knowledge of $A^{(m)}_N$ and
$E^{(m)}_n$ in a hierarchical manner.

\section{The Results}
The energy $E_n$ so obtained of the anharmonic oscillator given by
Eq.(1) to the fourth order in $\lambda$ is given by
\begin{eqnarray}
E_n[4]&=&\omega(n+\, \frac{1}{2})+\frac{\lambda}{2\omega}\, (n+\,
\frac{1}{2})-\frac{\lambda^2}{8\omega^3}\, (n+\,
\frac{1}{2})-\frac{\lambda^3}{\omega^4}\,
\Bigg[\frac{5}{96\omega}\, (2n+1) \nonumber \\&+&\frac{2}{3}\,
(4n^2+4n+1)\Bigg] +\frac{\lambda^4}{4\omega^4}\,
\Bigg[\frac{25}{96\omega^3}\, (n+\,
\frac{1}{2})+\frac{23}{12\omega^2}\, (4n^2+4n+1)\nonumber
\\&+&\frac{7}{2}\, (2n^2+2n+1)\bigg]+\ldots\, ,
\end{eqnarray}
\noindent The energy series given by (4) is a divergent-asymptotic
series, therefore one can use the Pad\'e approximants to calculate
the energy eigenvalues\cite{Loeffel}. The [N,M] Pad\'e approximant
to (4) is given by
\begin{eqnarray}
E[N,M]&=&E^{(0)}_n\, \frac{1+\lambda p_1+\lambda^2 p_2+ \ldots
+\lambda^M p_M}{1+\lambda q_1+\lambda^2 q_2+ \ldots +\lambda^N
q_N}\, , \nonumber \\&=&E^{(0)}_n+\lambda E^{(1)}_n+\lambda^2
E^{(2)}_n+ \ldots +\lambda^{N+M} E^{(N+M)}_n\, ,
\end{eqnarray}
\noindent where the coefficients $p_i(i=1, \ldots, M)$ and
$q_j(j=1, \ldots, N)$ in this equation can be calculated from the
knowledge of the energy coefficients $E^{(m)}_n$ up to the order
of $\lambda^{N+M}$. \\
\begin{table}[htbp]
\begin{center}
\begin{tabular}{c c c c c} \hline
$n$ & $\lambda$ & $E[4]$ & $E[3,3]$ & $E[3,4]$ \\ \hline
$0$ & $0.005$ & $0.501248$ & $0.501248$ & $0.501248$ \\
$$ & $0.01$ & $0.502493$ & $0.502493$ & $0.502493$ \\
$$ & $0.05$ & $0.512252$ & $0.512252$ & $0.512249$ \\
$$ & $0.1$ & $0.523620$ & $0.523634$ & $0.523590$ \\ \hline
$1$ & $0.005$ & $1.503740$ & $1.503740$ & $1.503740$ \\
$$ & $0.01$ & $1.507480$ & $1.507480$ & $1.507480$ \\
$$ & $0.05$ & $1.536260$ & $1.536260$ & $1.536240$ \\
$$ & $0.1$ & $1.566970$ & $1.567010$ & $1.566660$ \\ \hline
\end{tabular}
\end{center}
\caption{Energy eigenvalues as functions of the parameter
$\lambda$ for the $n=0$ and $n=1$ states.}
\end{table}

\noindent The numerical results are given in Tables 1, 2 and 3. In
the calculations we have the quadratic term in the potential as
one of the perturbation terms. In view of this, the energy
eigenvalues of the anharmonic oscillator are evaluated for
different values of the anharmonicity parameter $\lambda$ for the
eigenstates $n=0$ to 5. In the Tables, we also list the energy
eigenvalues E[4] which are correct to the fourth order of
$\lambda$.
\begin{table}[htbp]
\begin{center}
\begin{tabular}{c c c c c} \hline
$n$ & $\lambda$ & $E[4]$ & $E[3,3]$ & $E[3,4]$ \\ \hline
$2$ & $0.005$ & $2.506240$ & $2.506240$ & $2.506240$ \\
$$ & $0.01$ & $2.512450$ & $2.512450$ & $2.512450$ \\
$$ & $0.05$ & $2.559610$ & $2.559610$ & $2.559530$ \\
$$ & $0.1$ & $2.605020$ & $2.605080$ & $2.604100$ \\ \hline
$3$ & $0.005$ & $3.508730$ & $3.508730$ & $3.508730$ \\
$$ & $0.01$ & $3.517420$ & $3.517420$ & $3.517420$ \\
$$ & $0.05$ & $3.582290$ & $3.582290$ & $3.582120$ \\
$$ & $0.1$ & $3.637780$ & $3.637800$ & $3.635920$ \\ \hline
\end{tabular}
\end{center}
\caption{Energy eigenvalues as functions of the parameter
$\lambda$ for the $n=2$ and $n=3$ states.}
\end{table}

\begin{table}[htbp]
\begin{center}
\begin{tabular}{c c c c c} \hline
$n$ & $\lambda$ & $E[4]$ & $E[3,3]$ & $E[3,4]$ \\ \hline
$4$ & $0.005$ & $4.511230$ & $4.511230$ & $4.511230$ \\
$$ & $0.01$ & $4.522390$ & $4.522390$ & $4.522390$ \\
$$ & $0.05$ & $4.604310$ & $4.604310$ & $4.604000$ \\
$$ & $0.1$ & $4.665230$ & $4.665130$ & $4.662200$ \\ \hline
$5$ & $0.005$ & $5.513720$ & $5.513720$ & $5.513720$ \\
$$ & $0.01$ & $5.527350$ & $5.527350$ & $5.525180$ \\
$$ & $0.05$ & $5.625660$ & $5.625670$ & $5.378930$ \\
$$ & $0.1$ & $5.687380$ & $5.687040$ & $4.284910$ \\ \hline
\end{tabular}
\end{center}
\caption{Energy eigenvalues as functions of the parameter
$\lambda$ for the $n=4$ and $n=5$ states.}
\end{table}

\section{Conclusion}
Using the hypervirial relations (Eq.(3)), we have calculated the
energy coefficients $E^{(k)}_n$ of the anharmonic oscillator with
the potential given by Eq.(1) in a hierarchical manner. The energy
series is asymptotically divergent. We have then evaluated the
Pad\'e approximants E[3,3] and E[3,4] to the energy series. The
results for the potential without of the cubic term are in
agreement with those of Ref.[17]. Therefore, we conclude that the
hypervirial-Pad\'e summation method can be used to determine the
energy eigenvalues of the anharmonic potential given by Eq.(1). \\
The author thanks to Professor M. Onder for helpful discussions.

\end{document}